# Identifying and abating copper foil impurities to optimize graphene growth

N. Reckinger[1,*] and B. Hackens[1]

[1]IMCN/NAPS, Université catholique de Louvain, Chemin du Cyclotron 2, 1348 Louvain-la-Neuve, Belgium.

* Corresponding author: E-mail address: nicolas.reckinger@uclouvain.be (N. Reckinger).

# Abstract

Copper foil impurities are hampering scalable production of high-quality graphene by chemical vapor deposition (CVD). Here, we conduct a thorough study on the origin of these unavoidable contaminations at the surface of copper after the CVD process. We identify two distinct origins for the impurities. The first type is intrinsic impurities, originating from the manufacturing process of the copper foils, already present at the surface before any high-temperature treatment, or buried into the bulk of copper foils. The buried impurities diffuse towards the copper surface during high-temperature treatment and precipitate. The second source is external: silica contamination arising from the quartz tube that also precipitate on copper. The problem of the extrinsic silica contamination is readily solved upon using an adequate confinement the copper foil samples. The intrinsic impurities are much more difficult to remove since they appear spread in the whole foil. Nevertheless, electropolishing proves particularly efficient in drastically reducing the issue.

# Introduction

Chemical vapor deposition (CVD) on metallic catalysts is arguably the most promising method for graphene production at the industrial scale.[1] Owing to their easy handling, accessibility and low-cost, copper foils are the substrates of choice for graphene CVD growth. Since the CVD growth of graphene on copper is a surface-mediated process, it is substantially affected by the physical and chemical properties of the copper surface. Despite their huge popularity, copper foils present several downsides, related to their manufacturing process, adversely impacting the structural quality and the homogeneity of CVD graphene grown on top of them: uncontrolled crystallography, uneven topography (notably rolling striations) and impurities (at the surface and in the bulk). Tackling these issues is imperative for the reproducible synthesis of high-quality graphene on copper foils for future applications.

The control of the crystallographic orientation of copper foils has received considerable attention over the last years.[2,3,4,5,6,7,8,9,10,11,12] The preparation of smooth copper foils was also given wide consideration in the scientific literature. Methods such as electropolishing (EP)[13,14,15,16,17,18] and chemical mechanical polishing[19] can significantly reduce the roughness of copper foils. Alternatively, some research groups even produced their own copper foils to have a tighter control on their roughness by copper electrodeposition on a smooth, single-crystalline substrate and peeling-off.[20,21,22]

A few research groups analyzed the superficial chemical composition of copper foils from different suppliers, before or/and after graphene CVD growth. They all reported the presence of contaminants, regardless of the supplier or of the foil purity for a given supplier. They inferred that the impurities were of endogenous origin (segregating from the bulk of the copper foil during thermal treatment[23,24,25] or already present at the surface of the as-received foils[4,26,27,28]) or of exogenous origin (arising from the quartz tube[24,29,30,31,32,33,34] or back streaming from downstream system components[21]). Depending on their nature, these contaminants can catalyze the formation of graphene adlayers,[26,28] or also inhibit the growth of graphene, thereby creating holes[28] or dot defects in graphene,[34] or even etch graphene.[33] Impurities inside copper foils can also cause grain boundary pinning and prevent the growth of copper grains. Copper foils can also sometimes be coated with thin anticorrosion metallic oxide films that must be removed before CVD growth. [13,18,35,36]

The variety of the adverse effects of impurities in copper illustrates the need for reproducible methods, both to prepare impurity-free copper foils and also to prevent any contamination inside the CVD reactor during the growth itself. Many different pretreatment techniques have been proposed to prepare a clean and smooth copper surface. A first group of cleaning techniques involves a superficial chemical treatment in various liquids such as acetic acid,[35,14,15,16,37,38,39] solvents,[40] water,[14,41] inorganic acids (dilute $HNO_3$,[14,27] dilute HCl[27,42,43]), $FeCl_3$,[14,44] Cr or Ni etchants,[27] or a sequence of two chemicals.[26] They can be used to remove more or less efficiently the contaminants at the surface. However, they are useless against impurities located in the bulk of the foils that diffuse to the copper surface and precipitate during high-temperature (HT) processes, in particular during graphene growth. In contrast, the popular EP is more aggressive since it eliminates a non-negligible copper thickness from the foil, but it presents the joint advantage of removing surface and bulk contaminants and yielding a smooth surface. On the other hand, various methods were used to avoid the deposition of silica particles from the fused silica tube during CVD growth, either by working under a pressure three times higher than atmospheric pressure[29] or by confining the samples within a copper foil enclosure,[30] within an internal coaxial alumina screen tube and a tantalum sample holder[32] or within a heat-resisting box (whose precise nature was not mentioned by the authors).[34]

As already mentioned in 2015 by Wang *et al.*, nanoparticles contamination is frequently observed in postgrowth scanning electron microscopy (SEM) images of copper foils in the scientific literature. Today, despite the existing cleaning protocols, the simple examination of SEM or atomic force microscopy images found in a sample of recent papers[12,45,46,47] clearly reveals that the problem of contaminations is either not solved or is considered as secondary or even purely and simply swept under the carpet. In addition, many research groups understandably focus their attention mainly on the study of graphene and resort often only to optical microscopy or/and low/medium magnification SEM. In that case, the nanoscale details of the copper surface are not visible and impurity nanoparticles are thereby eluded. Nevertheless, considering its considerable impact on the CVD growth of graphene on copper foils, this is an important problem to solve if one is to draw reliable conclusion on intrinsic graphene nucleation mechanisms, kinetics, adlayer formation, *etc.* Overcoming this impurity-related lack of control is also a crucial step towards reaching standardized and industrial-scale graphene growth protocols.

In contrast with the very little attention paid in general by research papers to elucidate the nature and the origin of the copper foil contaminations in the framework of graphene growth, we have made it the central focus of the present article. Based on the existing body of literature, we conduct a systematic and thorough investigation. Our work aims to contribute to bridging the reproducibility gap faced by large-scale graphene synthesis.[48] For this purpose, we have observed and analyzed in detail the nature and the origin of the various types of contaminations that we systematically obtain after growing graphene on untreated copper foils. We have determined that the copper substrates were plagued by two distinct types of contaminants: (1) intrinsic impurities either already present at the surface of the copper foil under the form of micrometer-sized particles or impurities segregating from the copper bulk to the surface and precipitating after thermal treatment and (2) silica contamination originating from the furnace's quartz tube. In order to avoid the deposition of silica contamination, we have enclosed the copper substrates between two heat-resistant sapphire wafers separated by a small gap. In addition, EP was implemented to prepare a clean and smooth copper surface before CVD growth. We observe a significant reduction of the contaminants on copper after CVD, but without reaching total elimination. We believe that the present efforts towards contaminant characterization, and the strategy that we describe to avoid them, will provide a firm basis for more reproducible and standardized high-quality graphene growth.

## Experimental

*Preparation of the copper substrates*

**Electropolishing.** Copper foil pieces with a size of 2 × 2.5 cm$^2$ (cut from a 15 × 15 cm$^2$ foil, Advent Research Materials, 50 µm-thick, purity 99.9%) are used as starting substrates. The adopted EP procedure is inspired from Miseikis *et al.*,[49] with several adaptations. The homemade EP setup comprises a Schiefferdecker glass staining trough and one (or two) copper (1-mm-thick) electrodes (7.5 × 4 cm$^2$) for the cathode (see Fig. S1(a)). Before EP, we immerse the copper piece in acetic acid to remove the superficial copper oxide layer and get a clean copper surface. It is then connected to the power source, as the anode, via an electric cable terminated with a copper alligator clip. The cable is fixed to a stand. We only polish a surface of 2 × 2 cm$^2$ since the top of the copper piece where the alligator clip is attached is not immersed in the EP solution. After de-oxidation, the copper piece is rinsed in running deionized (DI) water. Next, the cathode is placed against the wall of the trough opposite to the anode which is placed at 1 cm from the wall. It is very important to keep the same geometry for obtaining reproducible results. After positioning the electrodes and connecting them to the power source, we add the electrolyte solution, composed of a 10:5:5:1 mixture of DI water, phosphoric acid (87%), ethanol and isopropanol, plus 16 g/l of urea per liter of DI water. We then apply a 9 V voltage to the anode, resulting in a ~1 A current for 1 min. When EP is finished, the copper piece is immediately rinsed in running DI water for 5 min, kept in vertical position and still attached to the alligator clip. This ensures a fast transfer to DI water after polishing in order to produce a clean copper surface without polishing residues. The copper piece is further rinsed with isopropanol and stored in the same liquid to avoid oxidation in air. Figure S1(b) illustrates a typical copper foil piece after EP.

**Electron beam evaporation.** The copper thin films were deposited on sapphire or copper foils at room temperature (at a pressure of ~$2 \times 10^{-7}$ mbar) by evaporation of 99.999% pure copper pellets from Kurt J. Lesker.

*Graphene CVD growth*

The CVD reactor is a horizontal hot-wall split furnace (MTI corporation) fitted with a quartz tube (1.2 meter-long, 4-inch outer diameter). The copper sample is first put on the unpolished side of a sapphire wafer (C-plane (0001), 3-inch diameter, 600-μm-thick, and single-side polished) to avoid Cu to be stuck on it, itself deposited on a quartz boat and inserted into the horizontal hot-wall split furnace's quartz tube at room temperature. After sealing the tube, it is pumped down to a pressure below $2 \times 10^{-2}$ mbar. An argon flow is then fed into the tube until atmospheric pressure is restored. This preliminary purge ensures that the oxygen concentration in the tube is well controlled (equal to the residual oxygen in the argon canister, *i.e.* 1 ppm).[50] The CVD procedure is directly inspired and slightly adapted from our previous works performed in a different CVD equipment.[4,36] This testifies to the transferability and interoperability of our growth process.[51] The temperature-time profile is presented in Fig. S2. First, the temperature of the furnace is increased from room temperature to 1050 °C during 1 h under a 300 sccm flow of argon and kept at 1050 °C for an additional duration of 30 min. Next, 200 sccm of $Ar/H_2$ (10% $H_2$ in Ar) are introduced for 20 min. Just after, dilute methane (200 ppm in Ar) at a flow rate of 30-40 sccm is injected during 1 h for graphene growth. Finally, the power supply is shut down and the furnace is left to cool down naturally in the same gas mixture. Note that, in the text below, annealing designs the exact same process but without injecting methane in the tube.

*Preparation of the copper foil cross-section*

Since the copper foils are very flexible, the copper foil piece is first sandwiched between two rigid 1-mm-thick copper plates glued with cold mounting epoxy (EpoFix from Struers). The copper foil piece was cut so as to comprise an electropolished zone and an unpolished zone (see Fig. S3(a)). The sandwich is left to dry for 24 h in air at room temperature. It is then included vertically in a flat, cylindrical mold filled with epoxy and again left to dry for 24 h. A dicing saw is used to cut the obtained cylinder in the middle, in a plane parallel to its circular faces (see Fig. S3(b)). One of the two halves is then used for the polishing procedure. The following steps were performed to prepare the copper foil cross-section: (1) grinding with SiC paper (P1200 grit), (2) polishing with diamond abrasive of 9 μm and then 6 μm in size, and (3) finishing step with 50-nm silica colloids. For SEM imaging, it was next covered with a thin conductive carbon layer. In order to determine correctly the copper etch rate, the backside of the electropolished copper foil piece used for the cross-section was protected by blue tape to prevent EP (see Fig. S1(c)). If not, copper will be etched slightly as well, even if not as much as the front side.

*Characterization techniques*

The SEM images were recorded with a Zeiss Ultra 55 microscope in in-lens detector mode at an electron energy of 2 keV in most cases. Energy-dispersive x-ray spectroscopy (EDX) was performed in another Zeiss Ultra 55 fitted with a Quantax Bruker system (with a silicon drift detector detector) at 5 or 15 keV. The optical microscopy images were obtained with a Zeiss Axio Imager Vario.

## Results and discussion

To facilitate the reading of the present paper, Fig. 1 schematically summarizes all of the experiments conducted during our investigation. We start our investigation with the common observation that, after heating at HT (see the details in the Experimental section), as-received (untreated) copper foil surfaces inspected by SEM at sufficiently high magnification are systematically decorated with micro- and nanoparticles standing out from the surface (see Figs. 2 and S4, the latter summarizing in a single SEM image all the adverse impacts of impurities on graphene growth). Noteworthy, micrometer-sized particles can already be found at the surface of as-received copper foils (see Fig. 2(a)). After HT treatment, such particles remain essentially intact but spread radially, forming a circular halo around them (see Fig. 2(b)). Such circular halos with a microparticle in the center are observed on all the inspected copper pieces after HT annealing, still in low concentration. By means of EDX, we have determined that the particle shown in Fig. 2(b) is composed mostly of aluminum oxide ($AlO_x$), with a small amount of magnesium (see Figs. S5(a) and (b)). On a total of ten such microparticles analyzed by EDX on eight different samples, three were composed of silicon oxide ($SiO_x$), six of $AlO_x$ (two with some magnesium) and one was a mixture of both (for comparison, see Table S1 summarizing reported elemental compositions of impurities on as-received copper foils). Since our work is performed in cleanrooms where silicon wafers are manipulated (most probably a very common situation in graphene-oriented laboratories), one could argue that the $SiO_x$ particles could come from the environment. However, considering the care taken in avoiding any cross contamination, this possibility is very unlikely. On the other hand, this argument can certainly be dismissed in the case of $AlO_x$-bearing particles. One could therefore reasonably assert that these microparticles were already present at the surface of copper and result from their manufacturing process.

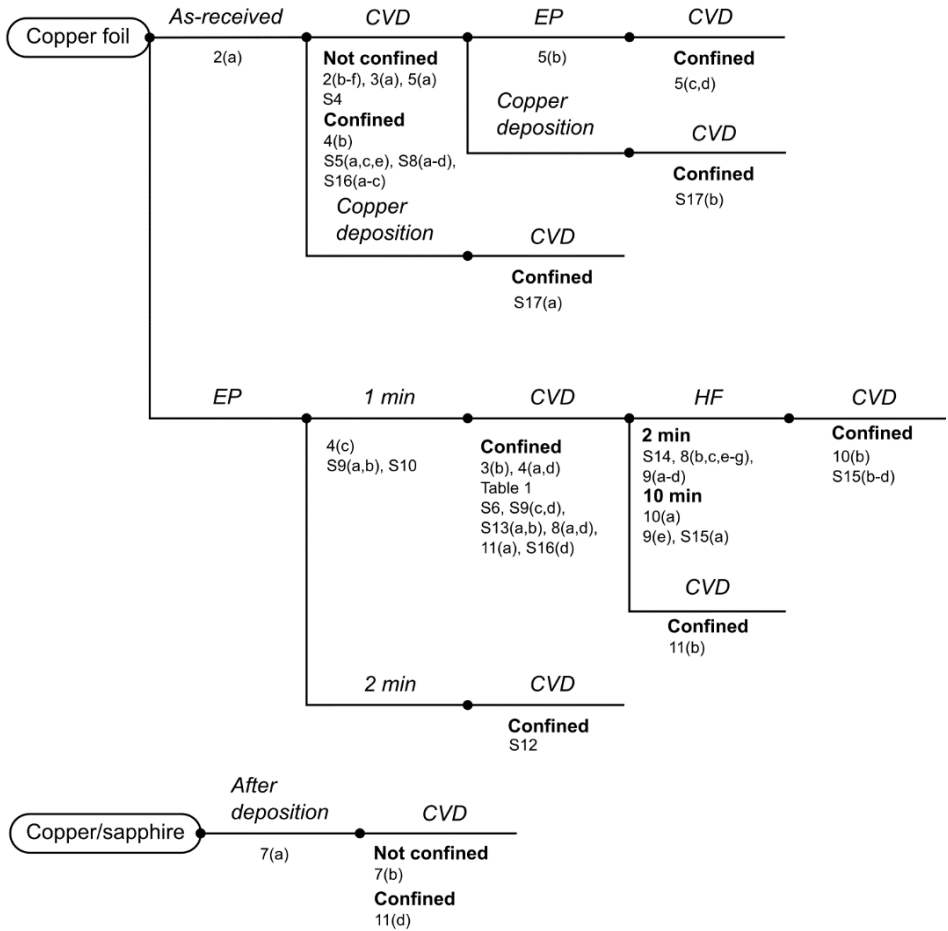

**Figure 1:** Schematic representation of all the experiments performed in this work. All the figure references are also associated to the corresponding process step.

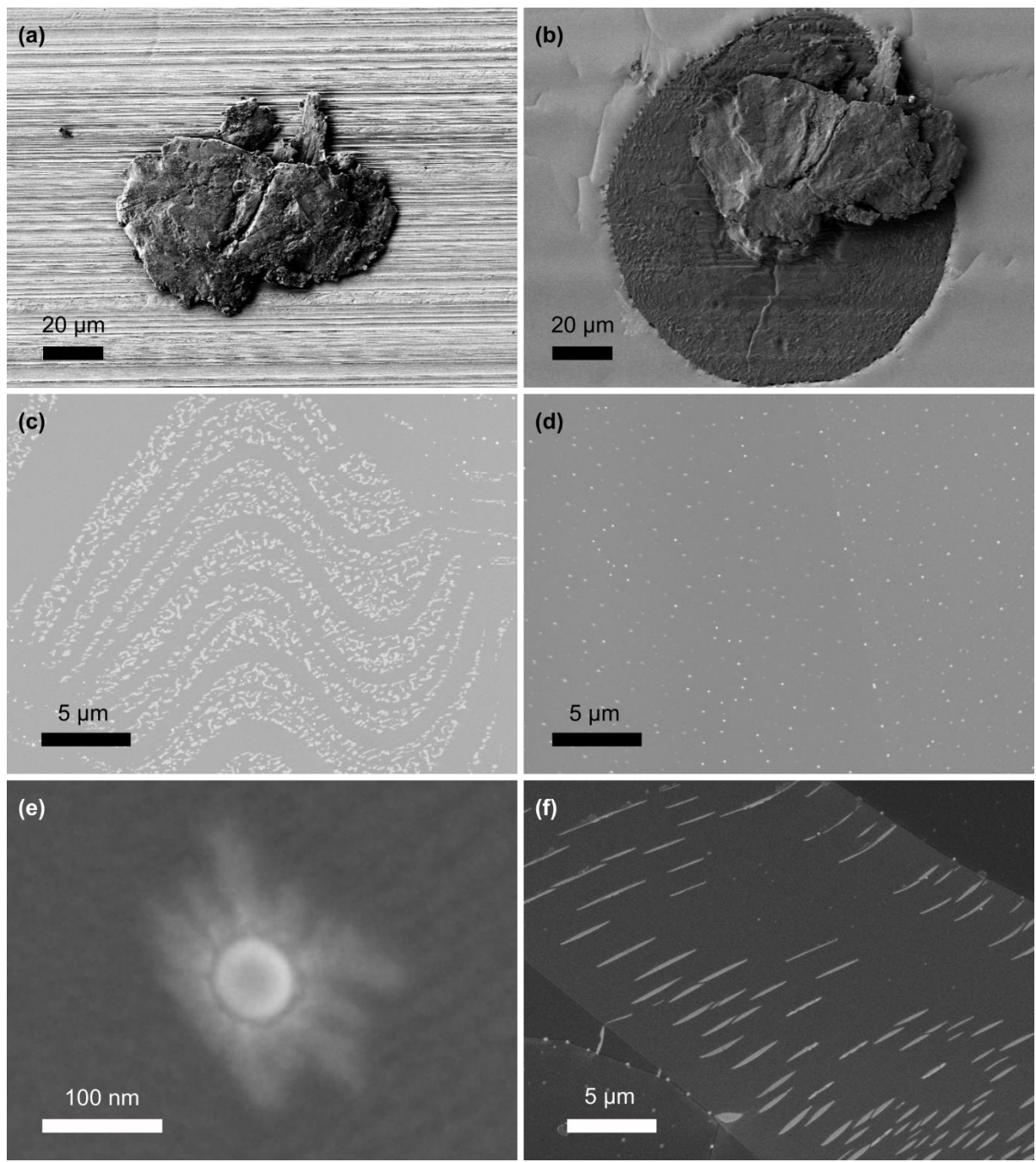

**Figure 2:** SEM images of (a) a microparticle on an as-received copper foil piece, (b) the same microparticle after HT treatment, (c) grey patches, (d) bright-contrasted nanoparticles, (e) a close-up view of such a nanoparticle, and (f) similar contaminations on copper foils from another manufacturer (Alfa Aesar).

Another type of contamination observed in SEM images of all the samples appears as sinuous, zebra-striped or ocellated patches displaying a light grey contrast (named "grey patches") (see Fig. 2(c)). These contaminations are strongly correlated to the morphology of the copper surface and their appearance and shape seem to depend on the crystallographic texture of the copper surface. They are too thin to be analyzed directly by EDX. Nevertheless, they are often found in the vicinity of thicker grey patches, yielding specific signatures in the EDX signal (see Figs. S5(c) and (d)). The EDX analysis shows that these sinuous structures are composed of $AlO_x$. Nanoparticles displaying a bright SEM contrast are also present all over the copper surface, either in association with the grey patches or isolated (see Fig. 2(d)). They are often spotted along topographic features of the copper surface such as grain boundaries and rolling striations either because it is energetically favorable to settle there or because it provides an easier segregation path (see Fig. S6). If one examines one of these nanoparticles by SEM at high magnification, one can observe that they comprise a central region (a "lenticular nanoparticle") surrounded by a greyish dendritic halo or corona (see Fig. 2(e)). Since they are quite small and thin, they are challenging to analyze by EDX. Nevertheless, by reducing the electron-beam energy to 5 keV (instead of the usual 15 keV) and using point-mode analysis, the composition of three of these lenticular nanoparticles could be determined to be a mixture of $AlO_x$ (mostly) and $SiO_x$ (see Figs. S5(e) and (f)), meaning that they are likely of endogenous origin (as discussed above). We have tried to identify the exact chemical nature of the halo by EDX as well under the same conditions, but to no avail probably because it is much too thin to give an exploitable signal. Note that the same type of contaminations (grey patches and lenticular nanoparticles with a halo) can be found on the popular 25-µm-thick Alfa Aesar copper foils after CVD as well (see Fig. 2(f)).

Previous literature has reported contaminations arising from the quartz tube.[32] The authors showed that copper evaporated from the copper foil at low pressure (LP) and HT can diffuse into the tube's quartz walls, causing its devitrification and the emission of SiO vapors. These vapors can then condense as $SiO_x$ nanoparticles on copper and on the growing graphene. In the present work, even though we perform the CVD process at atmospheric pressure, our equipment is also used for growing graphene by LPCVD. Therefore, it cannot be excluded that our samples suffer from the same type of contamination. To check for this, a simple confinement structure (see Fig. S7) was built by stacking two 3-inch sapphire wafers (see the experimental section) onto each other and separating them by means of four 3-mm-high high-purity copper cylinders (the copper pellets used for electron beam evaporation, see the experimental section). One as-received copper foil piece was placed between the two sapphire wafers, at the center, and another one was placed over the top sapphire wafer, left unprotected.

A striking difference showed up after observing the surface of the two samples by SEM after CVD growth: both samples were covered with lenticular nanoparticles, but these nanoparticles were surrounded by a dendritic halo only in the case of the unprotected sample (see Fig. 3(a)) and not for the covered one (see Fig. 3(b)). Moreover, interestingly enough, dendritic halos appear exclusively around lenticular nanoparticles sitting on copper and not if incorporated in graphene (see Fig. 3(a)). In Fig. 3(a), we can also notice that, as mentioned before, the lenticular nanoparticles can be found along copper grain boundaries. Based on the previous observations, we hypothesize that the lenticular nanoparticles serve as nuclei for the formation of the dendritic coronae. Since the halos nucleate exclusively on copper, they should be composed of a material showing a strong affinity for copper compared to graphene. It is plausible that they are made up of $SiO_x$ (or possibly $CuSi_xO_y$) condensing from SiO in vapor phase during the cooling step. In addition, it is worth noting that, contrary to a complex enclosure,[32] a simple stack such as ours is sufficient to prevent the condensation of the dendritic corona. This difference may be ascribed to the LP conditions used by the authors. Shivayogimath et al.[21] have also reported very similar dendritic contaminations that they have attributed instead to fluorocarbons backstreaming from the oil pump of the vacuum system. We believe this possibility to be less likely since our CVD system is fitted with a dry pump and is operated at atmospheric pressure during most of the CVD process. We have also performed another CVD growth to probe the size of the zone of exclusion by placing four samples arranged in a cross (see Fig. S7(a)) inside the surface delimited by the copper pillars (~4×5 cm$^2$). The conclusion is identical: the copper surface remains free of dendritic contamination in that area (see Fig. S8).

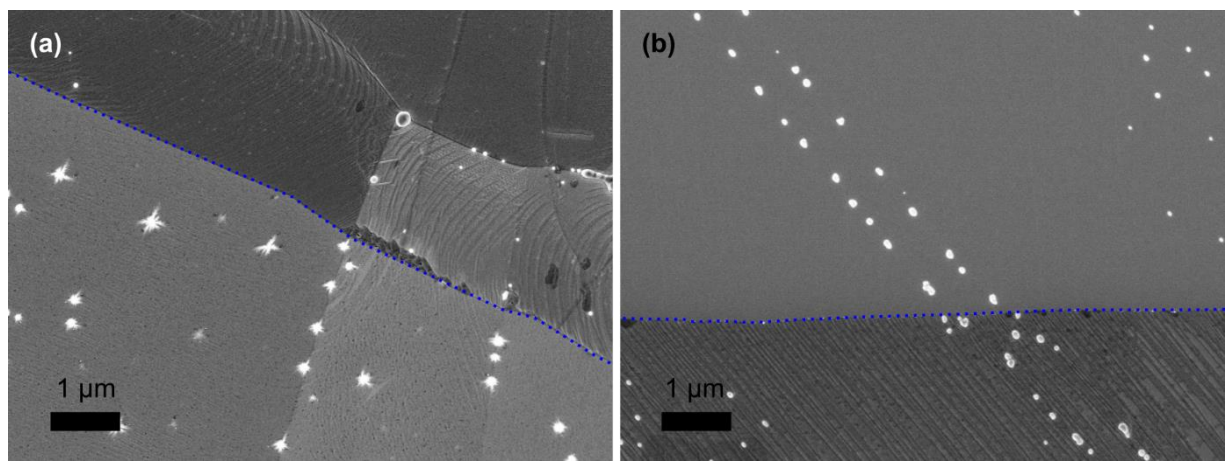

**Figure 3:** (a) SEM image of the surface of a copper sample piece grown (a) without or (b) with protection. Above (respectively, below) the dotted blue line, the surface is covered with graphene. Dendritic halos are observed only in the case of the unprotected sample.

After identifying and finding a way to eliminate a first extraneous source of contamination, we pursue our study by focusing on the necessary pre-treatment to remove the impurity particles inevitably present on the as-received copper foils. Among all the existing cleaning techniques as described in the introductory part, we choose to adopt EP in order to, at the same time, clean and make the copper surface smoother. First, Figs. 4(a) and (b) demonstrate, at a macroscopic scale, the efficiency of EP in terms of contamination inhibition where a polished and an unpolished area of the same copper foil piece after HT annealing are compared by SEM at low magnification (the inset to Figs. 4(a) and (b) shows a photograph of the corresponding copper foil piece). It can be clearly seen that the grey patches are no more visible in the electropolished zone. Then, by means of SEM at higher magnification, Figs. 4(c) and (d) compare the surface of a copper foil piece after EP and after CVD graphene growth at the same place (see Fig. S9 for the proof), respectively. At the same magnification, the inset to Figs. 4(c) and (d) illustrates the rough surface of an as-received copper foil, exhibiting rolling striations and defects. By comparing Fig. 4(c) and the inset, we can see plainly that EP is playing its part by smoothing out the rolling striations and thereby strongly reducing the average roughness of the copper foil piece. As a consequence, the copper grains become clearly apparent. Moreover, no contamination particles can be seen on the flat, electropolished copper surface, while it also appears covered with shallow cavities with average diameter around 100 nm (see Fig. 4(c)). More often than not, these recesses contain nanoscale particles (possibly contamination) but they escape detection by EDX because of their very small size (see Fig. S10). As seen in Fig. 4(d), after CVD, the copper surface is reconstructed (grain growth), is very smooth, but is nevertheless still covered with a few lenticular nanoparticles of the same type as mentioned previously.

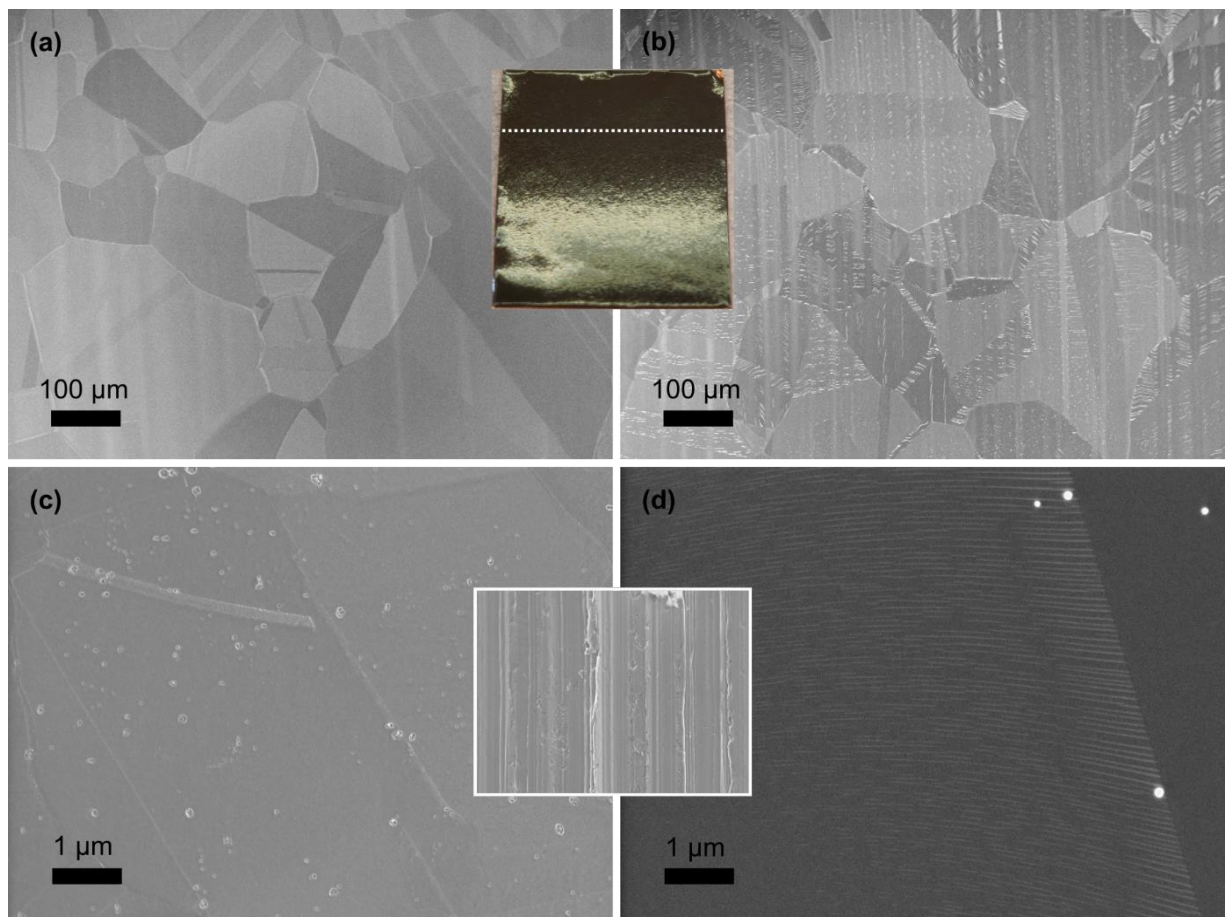

**Figure 4:** Low-magnification SEM image of the surface an electropolished 2 × 2.5 cm$^2$ copper foil piece after HT treatment in the (a) electropolished and (b) unpolished areas. The inset shows a photograph of the same piece, the dotted white line indicating the separation between the electropolished (below) and unpolished (above) areas. (c) SEM image of the surface of a copper foil piece just after EP. (d) SEM image of the same spot after HT treatment. The inset illustrates the surface of an unpolished copper foil piece at the same magnification.

The previous observations suggest that impurities are preferentially buried relatively close to the surface of copper foils and can diffuse towards the surface during HT annealing. By preparing and inspecting the cross-section of an electropolished copper foil piece, we have determined that about five micrometers of copper are removed after the process, under our EP conditions (see Fig. S11). Since EP removes these superficial copper layers, the impurities included in it are simultaneously eliminated, although not completely (as testified by the presence of a few lenticular nanoparticles). EP leads nevertheless to a much cleaner copper surface after CVD. Finally, in order to further improve the cleanliness of the copper surface, we have also performed the same kind of experiments with a copper foil piece annealed at HT before EP. The underlying idea would be to first segregate the impurities (all of them, if possible) buried into the copper bulk to the surface and next to remove them by EP. From Figs. 5(a-c) showing SEM images at moderate magnification of the same area before EP (and after annealing), after EP and after CVD graphene growth, we can draw the same conclusion about EP drastically inhibiting the occurrence of contaminants at the surface of copper. Nevertheless, SEM inspection at higher magnification can reveal that lenticular nanoparticles are still present, even though the amount of contamination is much weaker after EP (see Fig. 5(d)). We have also verified that doubling the EP time (2 min) does not result in further reducing the amount of contamination after CVD growth (see Fig. S12). This indicates that impurities are in fact likely distributed over the entire thickness of copper foils, albeit in smaller concentration in the bulk than close to the surface.

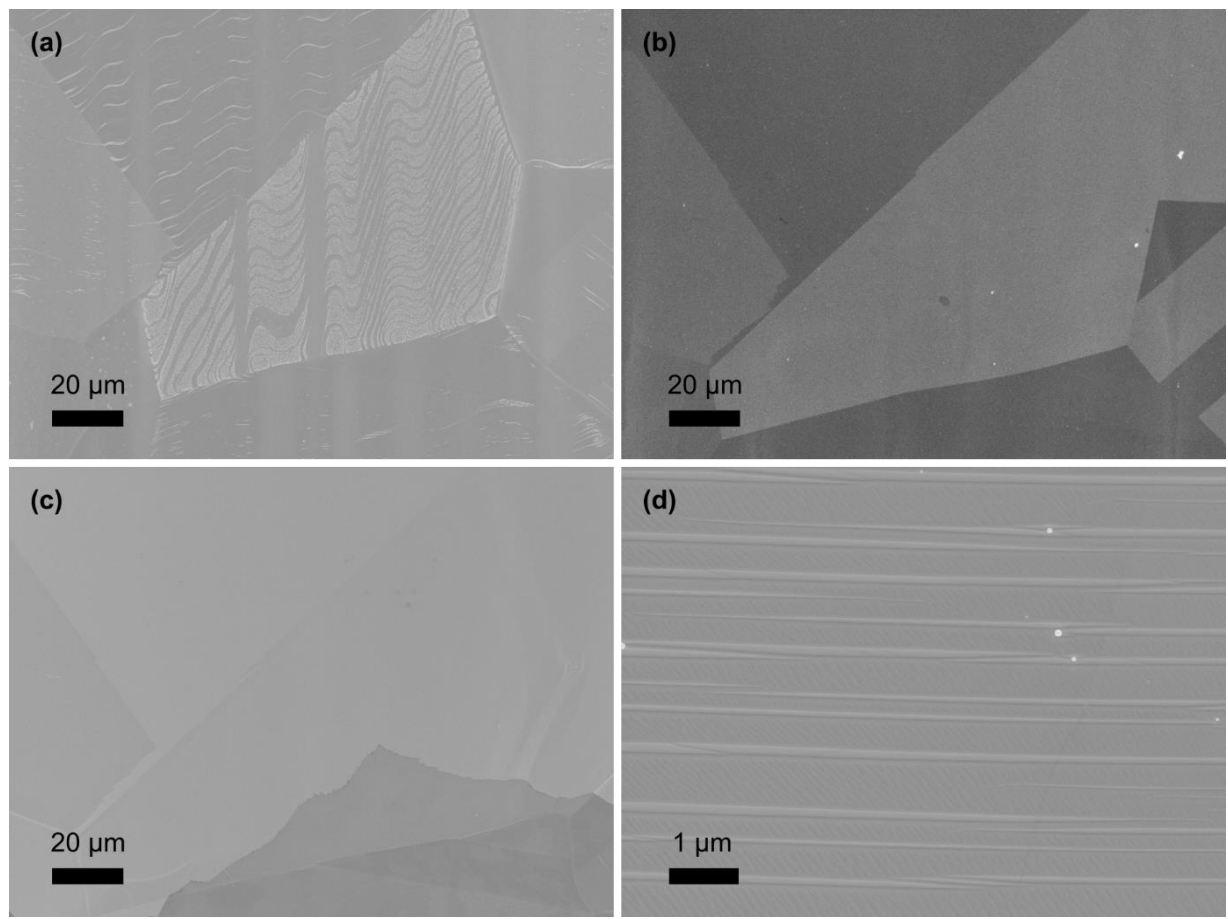

**Figure 5:** SEM image of the surface of a copper foil piece (a) after HT treatment, (b) after EP, and (c) after a second HT treatment, at the same place. (d) High magnification SEM view in the same condition as (c).

We now provide a quantitative evaluation of the amount of residual contamination nanoparticles after HT treatment of an electropolished copper foil piece. Since there may be some spatial variability inherent in the stochastic nature of diffusion processes and in the random distribution of impurities in the copper matrix, it is not sufficient to rely on a unique image. Therefore, we have analyzed SEM images in 70 random places all over the surface (see Fig. S13(a)), 20 at a ×3000 magnification (each image covering an area of ~1100 µm$^2$) and 50 at a ×10000 magnification (each image covering an area of ~100 µm$^2$) to compare the reliability of the data analysis as a function of the chosen magnification. For each magnification, Table 1 gives the average value (µ), the standard deviation (σ) and the coefficient of variation (σ/µ) of the number of nanoparticles per 100 µm$^2$, the relative nanoparticle surface density, and the mean nanoparticle size. Figure S13(b) illustrates the data extraction procedure, based on a particle identification algorithm in the Gwyddion software.[52] Figure 6 compares the histograms obtained for the ×3000 and ×10000 magnifications, respectively. In the ×10000 magnification case, the high coefficient of variation reflects the high spatial variability in terms of nanoparticle count per image (thus per 100 µm$^2$). As observed in Figure 6, it can indeed vary between 0 (for 7 images) and 199 (for one image). Note, however, that except for that very high value, all the other nanoparticle counts lie below 45 and the average value (µ = 11.5) is low. The surface occupied by the nanoparticles is also globally very low (on average, 0.012 µm$^2$) since the nanoparticles are themselves very small (41.6 nm on average). In the ×3000 magnification case, the trend is similar for the three extracted parameters. As with the ×10000 magnification case, the nanoparticle number was also significantly higher for one image. On average, the nanoparticle count per 100 µm$^2$ is nevertheless quite low, around 6. We can thus conclude that data analysis for the two magnifications yield similar results, provided we consider a high enough number of images. However, a difference between the two magnifications can be pointed out. The fact that 7 images present no nanoparticle for the ×10000 magnification can appear surprising at first sight. It can reflect the fact that the ×10000 magnification images are relatively small compared to the interparticle distance. Indeed, in the ×3000 magnification case, no spotless image could be found since their much larger size (11 times larger) makes that very improbable. In consequence, it could be an artifact related to the small image size. This illustrates the difficult compromise that must be made between the size of the image (it must not be too small to be statistically relevant) and the very small size of the nanoparticles (the image must be small enough to properly evidence the nanoparticles). This justifies precisely why this investigation was conducted at two magnifications. Moreover, it also indicates that the impurities are distributed very heterogeneously in the copper bulk, as testified by the high spatial variability noted above. The impurities possibly form clusters (as suggested by the two outlier values with a very high nanoparticle number in Figure 6) separated by cleaner areas. For reference, in Figs. S13(c) and (d), we plot the nanoparticle number per image for each magnification. We did not conduct the same analysis with an unpolished copper piece for comparison since, as visible in Fig. 5(a), the coverage in contamination can vary a great deal depending on the copper grain orientation.

| Magnification | ×10000 | | | ×3000 | | |
|---|---|---|---|---|---|---|
| | µ | σ | σ/µ | µ | σ | σ/µ |
| Nanoparticle number per 100 µm$^2$ | 11.5 | 28.7 | 2.5 | 6.4 | 11.9 | 1.9 |
| Relative nanoparticle surface density [%] | 0.012 | 0.015 | 1.22 | 0.018 | 0.019 | 1.05 |
| Mean nanoparticle size [nm] | 41.6 | 19.5 | 0.47 | 61.9 | 14.6 | 0.23 |

**Table 1:** Summary of the results obtained from the statistical analysis of nanoparticles at the surface of an electropolished and HT-treated copper foil piece.

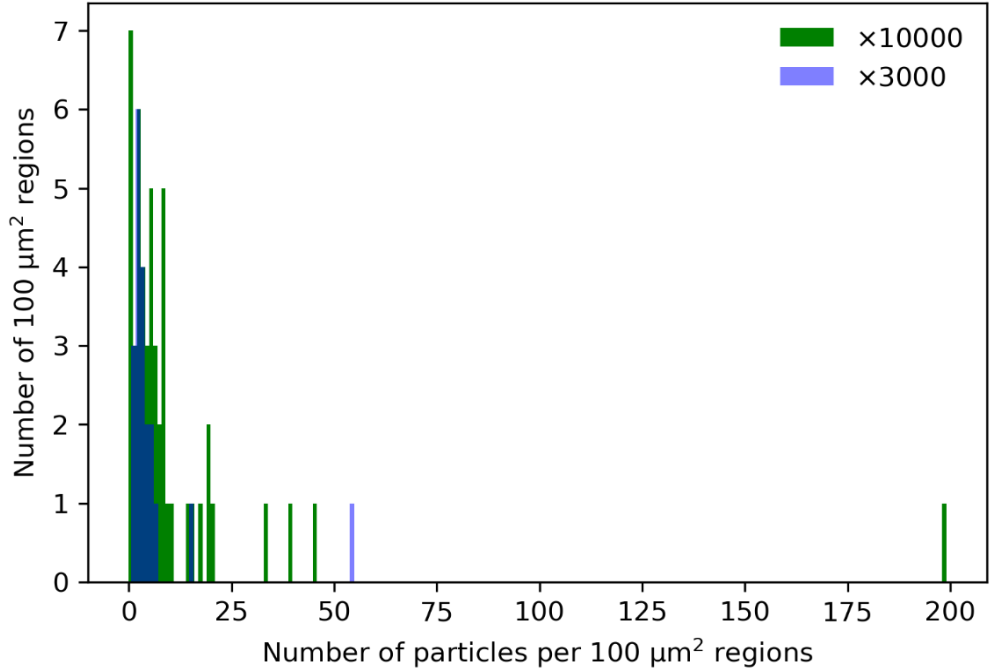

**Figure 6:** Histograms of the number of 100 μm² regions as a function of the number of nanoparticles per 100 μm² region for the ×3000 and ×10000 magnifications, respectively.

In order to consolidate our previous hypotheses on the origin of the impurities, we have conducted a series of additional control experiments. First, a 1-μm-thick highly pure copper film was deposited by electron beam evaporation on sapphire pieces cut from a 3-inch wafer (see the Experimental section). One copper-covered sapphire piece was confined between the two sapphire wafers separated by copper pillars and a second one was placed on the top sapphire wafer, as done before for copper foil pieces (see Fig. 3). As expected, after CVD, the copper surface of the covered copper/sapphire sample is much cleaner compared to its copper foil piece counterpart (see Fig. 7(a)) because of the huge difference in purity. Occasionally, we could find a few contamination nanoparticles possibly coming from the residual impurities contained in the copper source used for electron beam evaporation. Figure 7(b) shows a SEM image of the exposed sample. Interestingly, one can notice the presence of nanoscale contaminations reminiscent of the dendritic corona visible in Fig. 3(a). This confirms that this type of contamination nucleates on any impurity or asperity present on the copper surface and originates from a source exterior to the copper substrate. This also points out that even highly pure copper/sapphire substrates are not totally exempt from impurities.

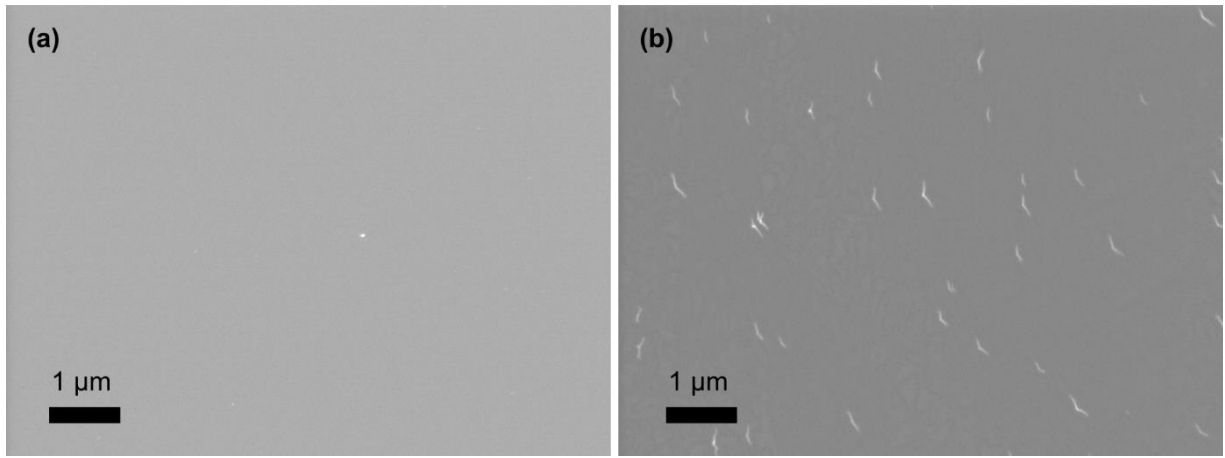

**Figure 7:** SEM image of the surface of a copper/sapphire substrate (a) with and (b) without protection, after HT treatment.

Secondly, considering that the contaminations are mostly composed of $SiO_x$ and $AlO_x$, we have investigated their removal with dilute hydrofluoric acid (2% HF) for a sample that was electropolished and HT-treated. We have observed that copper fluoride particles form on copper after exposing a slightly oxidized copper surface (the "native oxide" arising from conservation under air) to 2% HF (see Fig. S14). Consequently, we systematically de-oxidize the surface of copper with glacial acetic acid just before HF treatment. Next, we have determined the time required for entirely removing the lenticular particles. Figures 8(a) and (b) show that an etching time of 2 min removes most nanoparticles but not all of them. This may arise from slight differences in chemical composition between nanoparticles. The HF treatment also degrades the smoothness of the copper surface and the apparent roughness seems to vary slightly depending on the copper grain crystallographic orientation (see Fig. 8(c)). Note also that a small recess is left in copper after nanoparticle removal with dilute HF (see Figs. 8(d) and (e)). This image also illustrates that the simple fact of imaging the copper surface in a SEM modifies how dilute HF affects the copper morphology. Indeed, the surface of copper in Fig. 8(e) is not degraded while it should be. This can be explained by the fact that during the SEM scan, a thin carbon film is deposited on the surface which acts as a protective mask during exposure to dilute HF (see Figs. 8(f) and (g)).

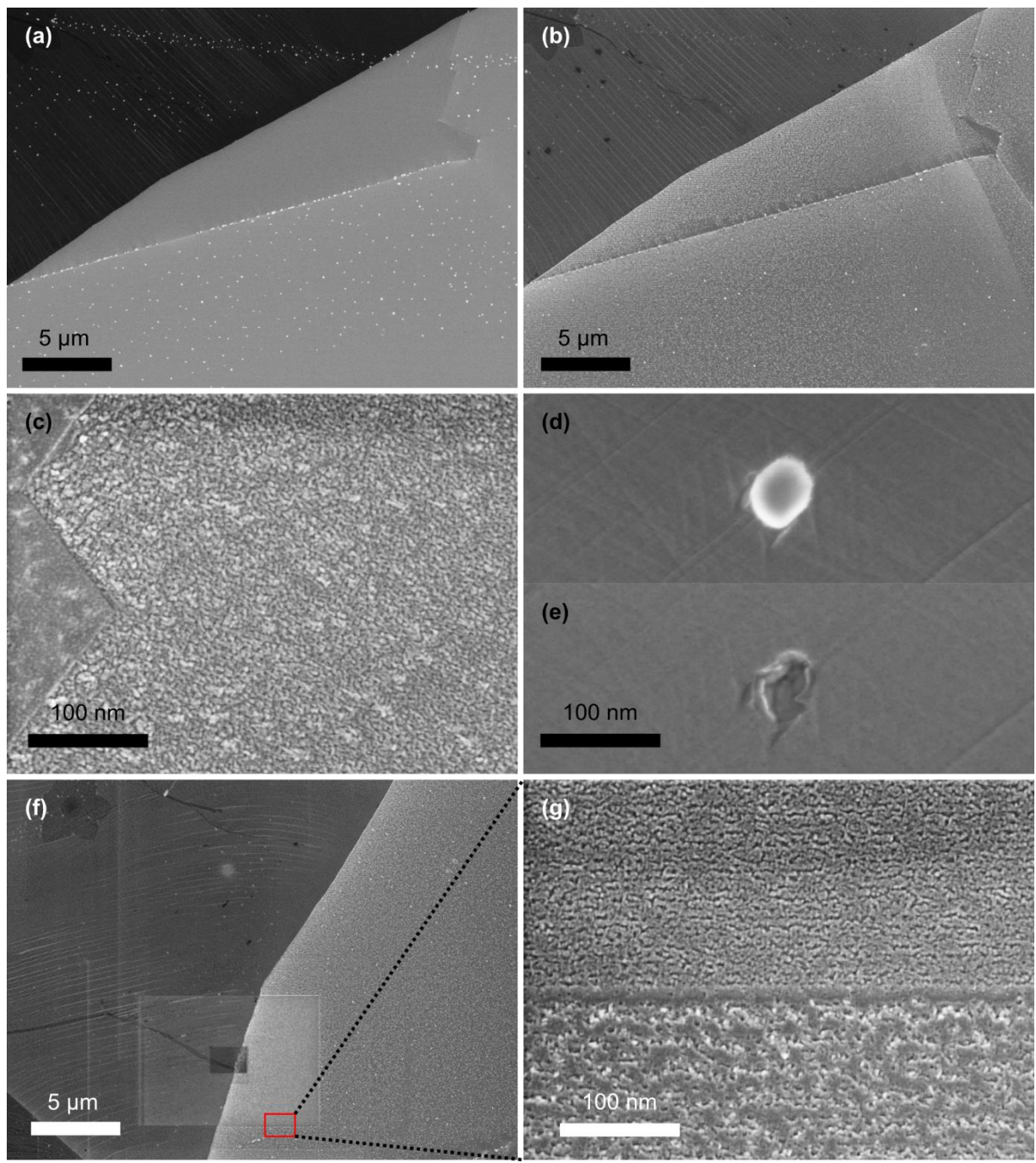

**Figure 8:** SEM images of the same region of a copper foil piece (after CVD growth) (a) before and (b) after treatment in dilute HF for 2 min to remove the contamination nanoparticles. (c) SEM image showing that the surface of copper is affected differently depending on the copper grain crystallographic orientation. Panels (d) and (e) compare the same lenticular nanoparticle before and after etching in dilute HF. (f) SEM image of an area of the same copper foil piece, after the dilute HF treatment. The area located at the bottom was observed by SEM before the treatment (rectangles resulting from scanning) and observed a second time in that image. (g) Zoom on the small region in the red rectangle in panel (f) illustrating how the copper surface is less affected if the area was imaged using the SEM before the dilute HF treatment because of a masking effect due to carbon deposition.

Compared to a bare copper surface, we have also noticed that many lenticular nanoparticles are more difficult to remove when embedded in graphene (see Figs. 9(a) and (b)). As seen in Fig. 9(c), larger contaminations causing holes in the graphene film are readily removed within a 2-min exposure to HF while lenticular nanoparticles may remain. Figure 9(d) even suggests that the nanoparticle visible in the image is covered by graphene, greatly complicating its access to any liquid. Indeed, it cannot be excluded that impurities continue to segregate to the copper surface even after graphene growth and that nanoparticles are thus sandwiched between copper and graphene. By increasing the immersion time in dilute HF to 10 min, the copper surface appears rougher but all the lenticular nanoparticles could be stripped away from the surface, both for bare and graphene-covered copper (see Figs. 10(a) and 9(e), respectively). We hypothesize that HF can reach the nanoparticle via nanoscopic defects in graphene. In Fig. 9(e), we can also spot the imprint left in copper by etched nanoparticles and the polygonal holes left in graphene due to graphene etching by hydrogen during CVD catalyzed by the same nanoparticles,[33] reflecting its hexagonal symmetry. The HF dip is thus a practical method to reveal defects in graphene. Even after cleaning with HF, Fig. 10(b) discloses the reoccurrence of lenticular nanoparticles after a second CVD growth. This confirms again that impurities are not depleted from the inside of the copper foil piece and that they can diffuse and precipitate at the surface after another CVD growth. Note that, in Fig. 10, the same area is inspected by SEM before and after CVD (see Fig. S15 for the proof).

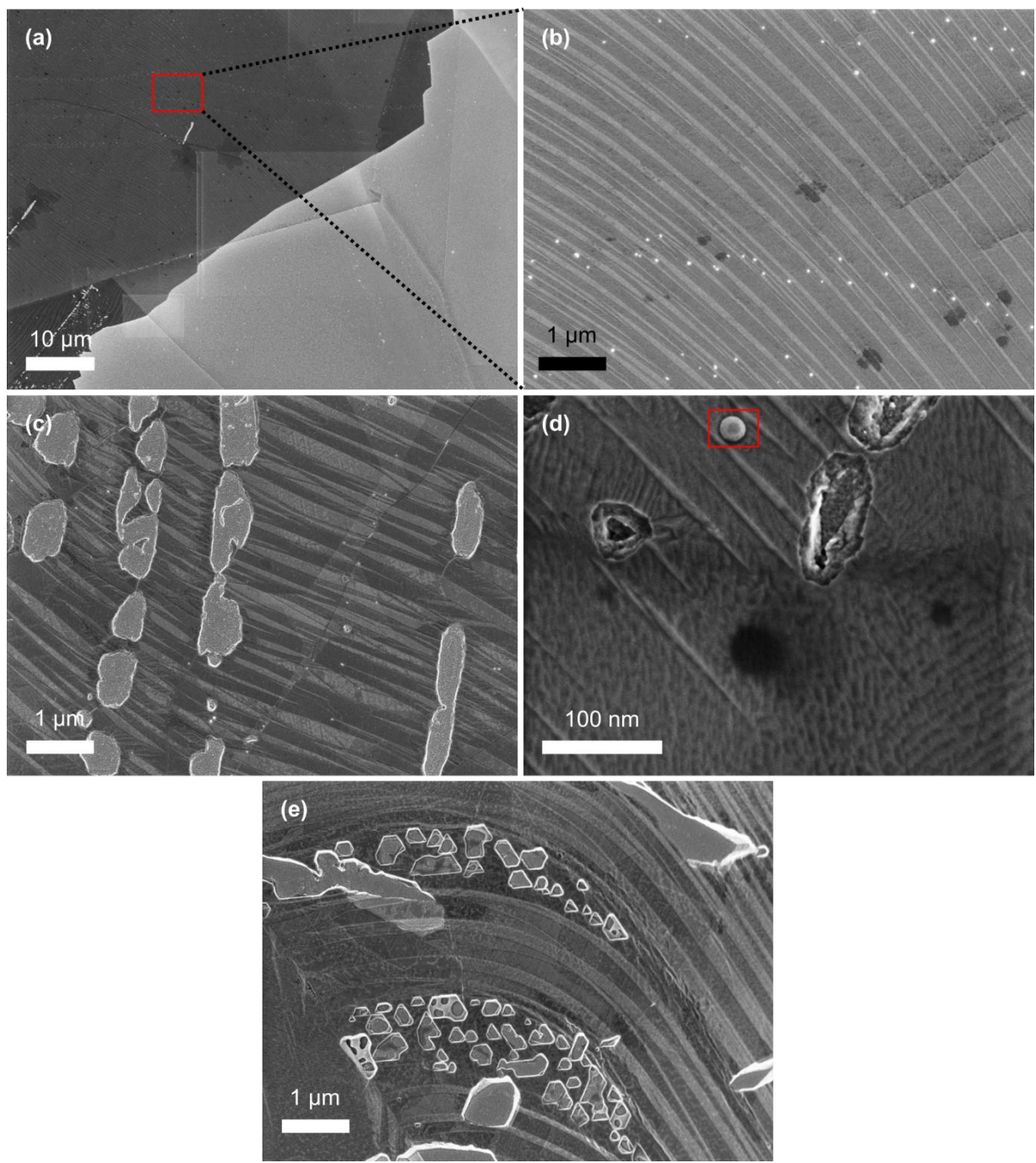

**Figure 9:** (a) SEM image of the same copper foil piece as in Fig. 8. The red rectangle delimits a zone (magnified in (b)) where nanoparticles seem embedded in graphene. Panel (c) shows a SEM image of another area of the same graphene flake where, in contrast, much larger particles have been removed. Panel (d) exemplifies a SEM image of a residual lenticular nanoparticle at high magnification. (e) SEM image showing the imprint left in copper by etched nanoparticles and the geometric holes left in graphene due to hydrogen etching of graphene during CVD by the same nanoparticles. This copper foil piece was treated in dilute HF for 10 min after CVD.

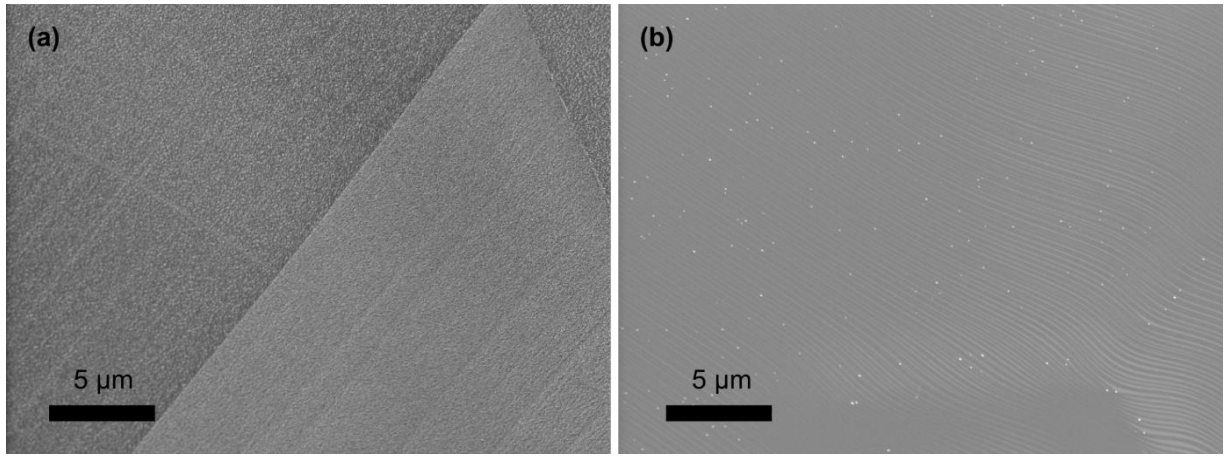

**Figure 10:** SEM image of the surface of a copper foil piece (a) after dilute HF treatment for 10 min and (b) after CVD growth, at the same place.

With the same idea in mind, we have also compared the same area of a copper foil piece after one CVD growth (see Fig. 11(a)) and after a second one (see Fig. 11(b)) with no intermediary treatment. From Fig. 11(b), it can be seen once again that more nanoparticles can precipitate at the surface. Note here that the copper grains were not reconstructed after the second CVD growth, greatly facilitating the identification, contrary to Fig. S15 for which the comparison procedure was more complicated. This difference can be explained by the fact that the copper piece in Fig. S15 was cut in smaller pieces before the second CVD growth. The simple fact of cutting the copper foil piece induces a new grain reconstruction in the sub-pieces after another CVD process. Moreover, it is also worth noting that this sample was completely enclosed by means of a clean quartz petri dish during the CVD process (see Fig. 11(c)), similar to the box technique.[34] Before, we have ensured that the quartz petri dish itself introduces no contamination by performing a CVD process with a copper/sapphire sample enclosed in the dish (see Fig. 11(d) as a proof). This definitively excludes an external source for the contaminations. In conclusion, this means that our HT treatment is too short to fully deplete the bulk of the copper foil from the impurities it contains.

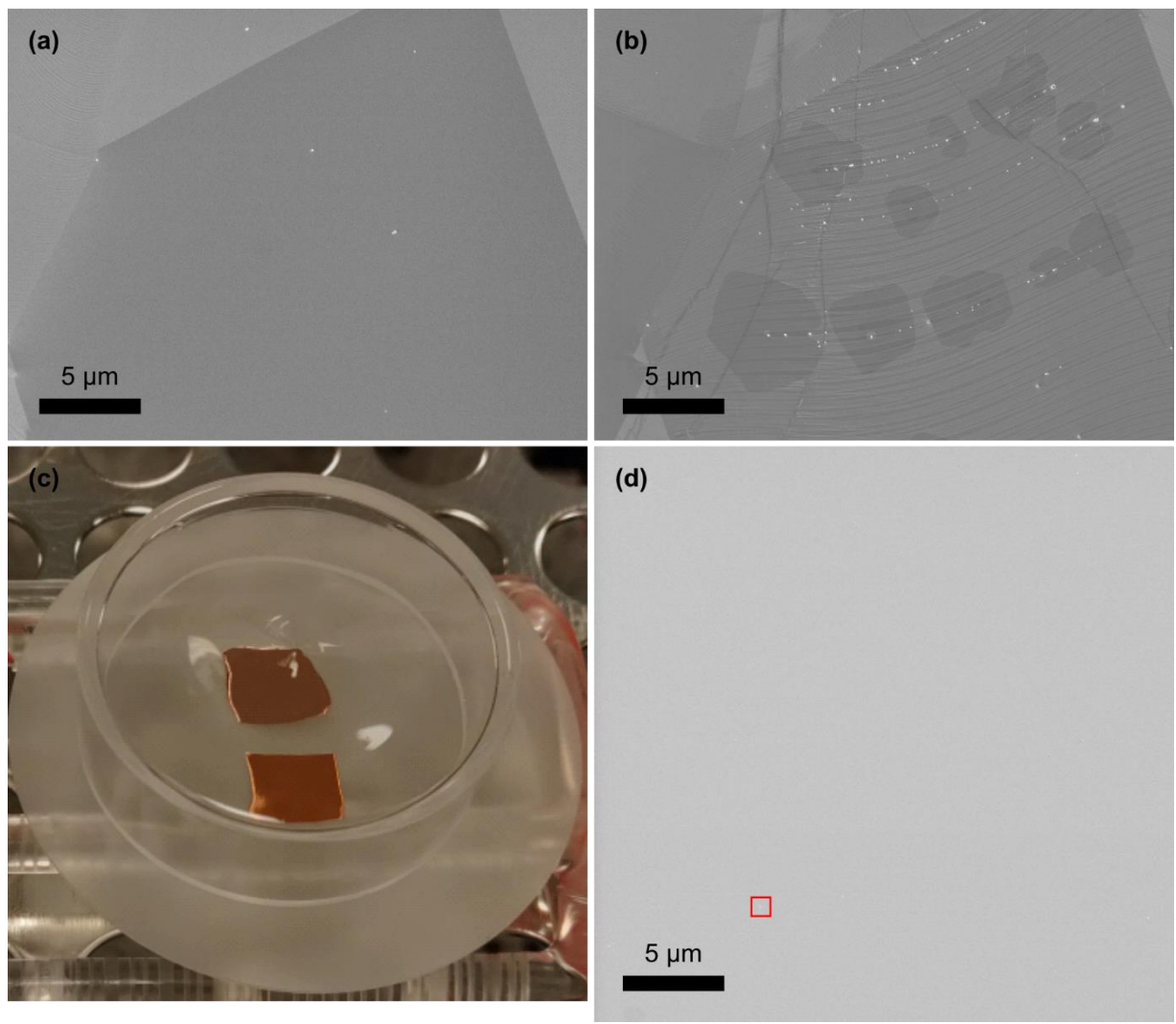

**Figure 11:** SEM images of a copper foil piece at the same spot before (a) and after (b) a second CVD growth. (c) Photograph of the quartz petri dish enclosing two copper foil pieces. (d) SEM image of a copper/sapphire sample, enclosed in the dish during the CVD process, after the CVD growth. The red rectangle comprises a nanoparticle to prove that the image is not out of focus.

It is worth discussing here the possibility that contamination may originate from lateral migration (for instance from improperly removed impurity particles or from the unpolished area (see the EP protocol) instead of segregation. Wang et al.[23] have indeed reported previously that impurity nanoparticles become mobile above 850 °C and could move over the duration of the CVD process (up to 250 µm in the straight-line, worst-case scenario for a CVD duration of 1h30). However, the authors also report that their movement is random, which strongly limits their travel distance. Figure 2(b) also confirms that nanoparticles arising from a micrometer-sized particle remain mostly concentrated around it, as testified by the circular halo of limited spatial extension. Consequently, in case a residual microparticle is left accidentally on copper following EP, the contamination will remain very local and it seems very unlikely that it may contaminate all the sample. We have also taken SEM images on both sides of the divide separating the unpolished and polished areas of a copper foil piece after CVD at a fixed abscissa and moving the sample down by steps of a few millimeters (see Fig. S16). The demarcation between the two areas appears clearly. This demonstrates that the unpolished area constitutes no "reservoir" of impurities that could contaminate the clean, polished zone and that if there were some lateral migration of impurities, it must be very limited. In all plausibility, we can therefore attribute the occurrence of the nanoparticles to segregation from the bulk rather than from drift at the surface of copper.

Based on the latter observation and with the hope of blocking the segregation of contaminants to the surface of copper, we have deposited by electron beam a 1-μm-thick high-purity copper film on two types of samples: an as-received copper foil piece and a copper foil piece covered by a monolayer of graphene. In both cases, the situation is the same: the high-purity copper layer is not sufficient to prohibit the segregation of impurities during CVD growth (see Fig. S17). It should not come as a surprise since, at the working temperature of 1050 °C, close to the copper fusion temperature (~1085 °C), the copper atoms are highly mobile and a copper layer cannot act as a stopping layer.

Even though the results of the previous experiments advocate strongly for impurity segregation from the copper bulk, they are still only indirect evidence of the presence of impurities in the interior of copper foils. In the next experiment, in order to provide a more direct proof, we have fully etched a copper foil piece deposited on a clean Si sample (see Fig. S18) in ammonium peroxydisulfate in order to collect any insoluble residues on the silicon surface. Prior to etching, the copper foil piece was immersed in 2% HF for 24 h to remove any $SiO_x$ and $AlO_x$ impurities on the copper surface. This way, we make sure that only impurities present in the copper bulk sediment on silicon. After etching, we have rinsed the silicon sample with water as gently as possible to avoid the impurities to be washed away. We have found residues showing up as aggregates of nanoparticles as well as micrometer-sized particles (see Fig. S19). In these aggregates, EDX analyses have evidenced the presence of metals like Cr, Sn and Fe (these metals being insoluble in ammonium persulfate) as well as elemental P and $SiO_x$. On the other hand, without surprise, the microparticles exhibited chemical compositions very similar to the two types of particles found at the surface on as-received foils, i.e. $AlO_x$ (with some Na) et $SiO_x$. In this way, we undoubtedly confirm the presence of impurities in the bulk of copper foils.

As a complementary experiment, with the hope of directly observing impurity microparticles, we have also investigated their presence in the copper foil cross-section. We could indeed find a few potential candidates, one of them could be identified by EDX as composed of $SiO_x$. However, considering the fact that the composition of that particle is not distinctive enough ($AlO_x$ would have been more irrefutable), that the polishing process itself could incur contamination (see the preparation protocol in the experimental section) and that large enough particles are rare (as inferred from our sedimentation experiment) and therefore difficult to find, it is preferable to remain inconclusive. It is worth noting that we have also observed nanoscale cavities in the copper foil cross-section (see Fig. S20), as already mentioned above at the copper surface after EP.

In the final experiment, a graphene sheet grown on an electropolished copper foil piece (with the top edge, where the alligator clip is attached, left unpolished, see Fig. S1(b)) is transferred onto a $SiO_2$ (285 nm)/Si substrate. Figures 12(a) and (b) exhibit two optical microscopy images of the electropolished and of the unpolished areas of that graphene sheet, respectively. The graphene film seen in Fig. 12(a) is continuous, mostly monolayer, with some adlayers inevitably present under such growth conditions (<5% coverage).[36] In Fig. S21(a), showing the same zone as Fig. 12(a) at lower magnification, a few shapeless regions (highlighted in red rectangles) exhibiting a darker contrast and a larger surface compared to typical adlayers can be observed occasionally. This type of regions (see a SEM view focusing on one of them in Fig. S21(b)), named branch-like multilayer graphene regions, were reported previously for graphene grown under similar conditions to ours, on the same type of copper foil and in the same CVD equipment.[50] They were attributed to the local accumulation of impurities in the vicinity of the copper surface during thermal annealing under residual oxygen. Another type of multilayer graphene region that could be observed once in a while are, as we believe, a linear version of the branch-like multilayer graphene regions, showing up as scars (see Fig. S21(c)). In stark contrast, on the unpolished zone (see Fig. 12(b)), the graphene film is covered with a much higher density of graphene adlayers, branch-like and scar-like multilayer graphene regions compared to the electropolished side. This observation is a supplementary testimony to the preponderant role of EP in considerably decreasing the quantity of contamination diffusing to the copper surface during HT processes.

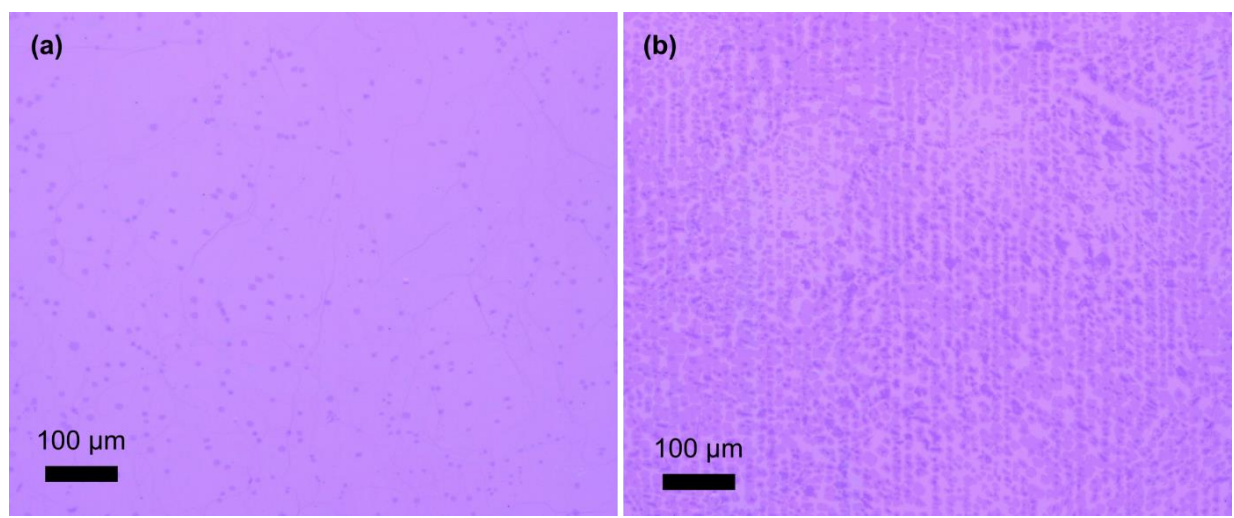

**Figure 12:** (a) Optical microscopy image of a graphene sheet synthesized on an electropolished copper foil piece and transferred onto a SiO$_2$/Si substrate. (b) The same graphene sheet grown in the small area left unpolished (top border) of the copper foil piece.

## Conclusion

Even though the investigated copper foils are pure at 99.9%, the remaining 0.1% impurities constitute a real nuisance in the context of reproducible and defect-free graphene growth. EP has not only proved useful for smoothening out the copper surface but also to remove the first copper micrometers where most impurities appear to reside. The amount of contamination after CVD growth was in consequence strongly reduced after EP, but not completely suppressed. Achieving full impurity removal is not possible with pre-growth superficial chemical treatments. Albeit indirectly or semi-indirectly, our bulk impurities segregation/precipitation scenario during HT treatment is supported by a large body of evidence. The present results could be further corroborated by dynamic, *in situ* studies[23] to go beyond the before/after comparison of SEM images and by analyzing in more details the chemical composition of nanoparticles by appropriate surface techniques such as scanning Auger microscopy.[53] As a final remark, the results of this study strongly advocate in favor of home-made copper foils by electrodeposition,[20,21,22] still very confidential nowadays.

## Acknowledgments

B. H. (Senior Research Associate) acknowledges financial support from the F.R.S.-FNRS (Belgium). The present research was funded by the Fédération Wallonie-Bruxelles through the ARC grant No. 21/26-116, by the Flag-Era JTC project "TATTOOS" (grant No. R.8010.19) and by the European Union's Horizon Europe research and innovation program under grant agreement No. 101099139 ("FLATS" project). The authors would also like to thank Laurence Ryelandt and the LACaMi platform for their technical and scientific support, as well as Benjamin Huet for useful discussions.